\documentstyle[aps,epsf,prc]{revtex}

\begin{document}
\title{ Photoproduction of charmed vector mesons, 
$\gamma+N\to  {\cal B}_c +\overline{D^*}$, with ${\cal B}_c=\Lambda_c$
or $\Sigma_c$ }
\author{Michail P. Rekalo \footnote{ Permanent address:
\it National Science Center KFTI, 310108 Kharkov, Ukraine}
}
\address{Middle East Technical University,
Physics Department, Ankara 06531, Turkey}
\author{Egle Tomasi-Gustafsson}
\address{\it DAPNIA/SPhN, CEA/Saclay, 91191 Gif-sur-Yvette Cedex,
France}
%\date{\today}

\maketitle
\begin{abstract}
The energy and the angular dependence of the associative vector charmed ${D^*}$-meson photoproduction, 
$\gamma+N\to {\cal B}_c+ \overline{D^*}$, with ${\cal B}_c=\Lambda_c$ or $\Sigma_c$, has been predicted in framework of pseudoscalar D-meson exchange model. The behavior of the cross section is driven by phenomenological form factors, which can be parametrized in terms of two independent parameters. The predicted values of the cross section are sizeable enough to be measured in the near threshold region.
\end{abstract}


%\section{Introduction}

It is well known that, due to the small mass difference between the vector $D^*(2.010)^\pm$ and $D^*(2.007)^0$, the pseudoscalar
$D^{\pm}$ and $D^{0}$ charmed mesons, the radiative decay 
$D^*\to D+\gamma$ is very important for the neutral $D^{*0}$ $(Br\simeq 40\%)$
and it is negligible for the charged $D^{*\pm}$ $(Br\simeq 1\%)$, in comparison to the pion decay $D^{*\pm}\to D+\pi$. Therefore, the absolute value of the transition magnetic moment (TMM) for the decay $D^*\to D+\gamma$ allows to determine the value of the radiative width, $\Gamma (D^*\to D\gamma$), for the neutral and charged vector mesons, and, consequently, to find the total widths of the vector  $D^*$-meson, which has not been determined up to now: 
$\Gamma _{tot}(D^{*\pm})\le 0.131$ MeV, $\Gamma _{tot}(D^{*0})\le 2.1$ MeV \cite{pdg}. These TMM, which are generally different for the quoted decays,
$D^{*\pm}\to D^{\pm}+\gamma$ and $D^{*0}\to D^{0}+\gamma$, are particularly interesting for testing the predictions of many theoretical approaches [2-12], such as, for example, quark model, dispersion sum rules or HQEPT. The knowledge of the TMM can also allow to predict the branching ratio for the conversion decay $D^{*}\to D+e^+ +e^-$ \cite{Al94}.

The standard method to determine the radiative widths $\Gamma (D^*\to D\gamma$) through the Primakoff effect, which has been successfully used for the decays $\rho\to \pi +\gamma$ or $ A\to\pi+\gamma$, can not be applied here, due to the short decay time of the $D$-meson ($\tau(D)\simeq 10^{-13}$ s \cite{pdg}). 
In order to study these TMM a unique way is represented by the associative photoproduction of charmed particles, for example, by the process 
$\gamma +p \to\Lambda_c(\Sigma_c) +\overline{D_c}$, near threshold. The large threshold  ($E_{\gamma}\simeq $9 GeV) will allow to investigate these reactions, after the upgrade of the electron accelerator at the Jefferson Laboratory (JLab) \cite{gen}.

The cross sections for such processes, in the near threshold region, are sizeable enough to be experimentally accessible \cite{Re01}. Moreover it was shown 
that the angular dependence of the differential cross section and the $\Sigma$- asymmetry (with linearly polarized photons), for the reaction 
$\gamma +p \to \Lambda_c ^++\overline{D_c^0}$, are sensitive to the value of the TMM of the $\Lambda_c$-hyperon.

Our aim here is to consider the processes of associative charm photoproduction of vector mesons and to study the sensitivity of the  differential and total cross sections to the TMM for $D^*\to D$ in the case of pseudoscalar D-exchange. The processes $\gamma+N\to \Lambda_c (\Sigma_c) +\overline{D^*}$ are the simplest two-body reactions of photoproduction of the charmed vector $ D^*(2010)$-meson on nucleon. Their threshold is high: $E_{\gamma}$= 9.361 (10.144) GeV, for $\Lambda_c (\Sigma_c)$-production. 

In order to discuss the reaction mechanism, for the reaction  $\gamma+N\to  {\cal B}_c +\overline{D^*}$( with ${\cal B}_c=\Lambda_c$ or $\Sigma_c$) in the near threshold region, we will proceed by analogy with other vector meson photoproduction processes, such as $ \gamma+N\to N+V,~V=\rho,~\omega,~\phi,$ and $\gamma+N\to K^*+\Lambda(\Sigma).$ In case of neutral vector meson photoproduction the diffractive mechanism dominates at large photon energies. In the near threshold region other mechanisms are more important: for  $\omega$-photoproduction  the one-pion exchange dominates, whereas in case of $\rho$-photoproduction,  due to the small $\rho\pi\gamma$
coupling constant, the scalar $\sigma$-exchange has to be considered \cite{Fr96}. Considering the coupling constant $g_{\rho \sigma\gamma}$ as a fitting parameter, it is possible to reproduce the $t-$ dependence of the differential cross section for the process $\gamma+p\to p+\rho^0$ - in the near threshold region. However a direct measurement of the $\rho\to \sigma^0\gamma$ decay \cite{Ac00} gives an 
experimental value of the  $g_{\rho \sigma\gamma}$-constant  smaller than the fitted value \cite{Fr96}, which is necessary to reproduce the absolute value of the differential cross section for the process $\gamma+p\to p+\rho^0$ at $E_{\gamma}\le 2$ GeV. One could still reproduce the data by increasing the value of the coupling constant for the $\sigma NN$-vertex.

In case of $\rho^{\pm}$ or $K^*$-meson photoproduction, the diffractive mechanism is forbidden, for any kinematical condition. This applies also to the process 
$\gamma+N\to  {\cal B}_c +\overline{D^*}$. For this last process we can conclude that the pseudoscalar $D-$exchange has to dominate, in analogy with the $\pi$ or $K$-exchange for
$\gamma+p\to p+\omega$ or $\gamma+p\to K^*+\Lambda(\Sigma)$.

In principle, in addition to meson exchange, other mechanisms can occur. For example,  nucleon resonances $N^*$ strongly contribute  in the resonance region, for strange particle or vector meson photoproduction \cite{Zh98}.  On the contrary, the threshold for the process $\gamma+N\to  {\cal B}_c +\overline{D^*}$ is so large, that there is no physical reason to include these processes. One might consider, for 
$\gamma+p\to \Lambda_c+\overline{D^*}$, the one-baryon exchange in $s$ and $u$-channel as it has been done in \cite{Re01} for the photoproduction of pseudoscalar D-meson, $\gamma+N\to  {\cal B}_c +\overline{D}$. However we will not include these two diagrams in our model. The reason is that these contributions contain at least two unknown coupling constants, in the vertex 
$N {\cal B}_c \overline{D^*}$ (corresponding to the Dirac (vector) and Pauli (tensor) interactions), decreasing essentially the predictive power of the model. Moreover, the previous experience with similar diagrams, in the case of the processes $\gamma+N\to N+ \rho(\omega)$ showed that these contributions are not essential in the differential and total cross sections.

%\section{Photoproduction of vector mesons, 
%$\gamma+N\to  {\cal B}_c +\overline{D^*}$, with ${\cal B}_c=\Lambda_c$
%or $\Sigma_c$}

The TMM for the $D^*\to D+\gamma$ decays determine the
matrix element for the exclusive process 
$\gamma+N\to  {\cal B}_c+ \overline{D^*}$, when  the pseudoscalar
$D$-exchange (Fig. \ref{fig:dst}) is considered in complete analogy with the
$\pi$-exchange for the process $\gamma+N\to N+V^0$, $V^0=\rho$, or $\omega$.
The pseudovector $D^*$-meson photoproduction seems preferable to the pseudoscalar $D$-meson photoproduction for the determination of
the coupling constant $g_{D^*D\gamma}$, because, in
the first case, there is only one strong coupling constant, $g_{N{\cal B}_cD}$ (instead of
$g_1$ and $g_2$ for the process $\gamma+N\to  {\cal B}_c +\overline{D}$).

The matrix element for $\gamma+N\to  {\cal B}_c+ \overline{D^*}$, in framework of $D$-exchange, can be written in the following form:
$${\cal M}(\gamma N\to {\cal B}_c\overline{D^*})=
\displaystyle\frac
{ie}{m_{D^*}}\displaystyle\frac {g_{D^* D \gamma}}{t-m^2_D}
{g_{N{\cal B}_cD}}
\overline{u}(p_2)\gamma_5{u}(p_1)\epsilon_{\alpha\beta\gamma\delta} 
e_\alpha k_{\beta} U_\gamma q_\delta ,$$
where $U_\gamma (e_\alpha)$ is the four-vector of the produced $D^*$-meson (initial $\gamma$) polarization, $U\cdot q=0~(e\cdot k=0)$, $k$ and $q$ are the four-momenta of $\gamma$ and $D^*$, $t=(k-q)^2$, $m_D~(m_D^*)$ is the mass of $D~(D^*)$-meson, and $g_{D^* D \gamma}~(g_{N{\cal B}_cD})$ is the coupling constant for the vertex $D^*\to D \gamma~(N\to{\cal B}_cD)$.

After summing over the polarizations of the final particles ($\overline{D^*}$ and ${\cal B}_c$) and averaging over the polarizations of the initial particles ($\gamma$ and $N$) one finds the following expression for the differential cross section of 
$\gamma+N\to  {\cal B}_c +\overline{D^*}$:
$$\displaystyle\frac{d\sigma}{d|t|}= \displaystyle\frac{\alpha}
{16} g^2_{N{\cal B}_cD} \displaystyle\frac{g^2_{D^* D \gamma}}{(s-m^2)^2} \displaystyle\frac{(M-m)^2-t}{m_{D^*}^2}
\left ( \displaystyle\frac{t-m_{D^*}^2}{t-m_D^2}\right )^2 F^2(t),$$
where $s=(k+p_1)^2$ and $M(m)$ is the ${\cal B}(N)$ mass.

Following the analogy with one-pion exchange, it is necessary to introduce in the expression of the cross section a $t$-dependent form factor $F(t)$, normalized to unity, for $t=m_D^2$:
\begin{equation}
F(t)=\displaystyle\frac{1}
{\left (1-\displaystyle\frac{t-m_D^2}{\Lambda^2}\right )^n},
\label{eq:eqff}
\end{equation}
where $n=1$ or 2 and $\Lambda$ is a cut-off parameter. 
Such form factors are necessary ingredients of this phenomenological model and are introduced in order to improve the $t-$behavior of the differential cross section, in the region of large values of $|t|$. For charm particle photoproduction the cut-off parameter $\Lambda$, in Eq. (\ref{eq:eqff}), 
is different from the case of light vector meson photoproduction, as 
it reflects the different energy scale: the typical size of charmed particles (meson and baryon as well) is essentially smaller than the one of hadrons or mesons: this leads to a value of $\Lambda\simeq m_{J/\psi}$. In numerical estimations we will use values of $\Lambda$ in the range between 2 and 3 GeV.

In the near threshold region, $t$ is not a good physical variable, as $|t_{min}(\cos\vartheta=1)|\simeq m_D^2\simeq$ 4 GeV$^2$, therefore  the  differential cross section ${d\sigma}/{d\Omega}$ is more preferable. This observable is reported in Fig. \ref{fig:difcs}, for two different form factors, $n=1$ and 2, considering two different values for the parameter $\Lambda$,  $\Lambda$ =2 and  3 GeV.
 
\begin{figure}
%\mbox{\epsfxsize=15.cm\leavevmode \epsffile{fig4.ps}}
\caption{Dependence of the backward-forward asymmetry for the processes  $\gamma+p\to \Lambda_c^+ +\overline{D^{*0}}$ (a) and $\gamma+p\to \Sigma_c^+ +\overline{D^{*0}}$ (b) at $E_\gamma=11$ GeV, for $n=1$ (solid line) and $n=2$ (dashed line).}
\label{fig:ratio}
\end{figure}
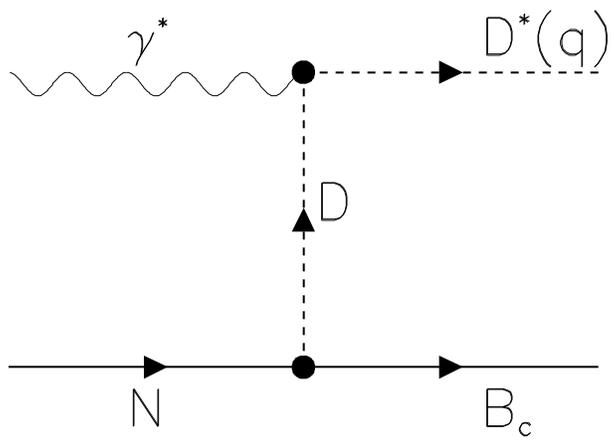
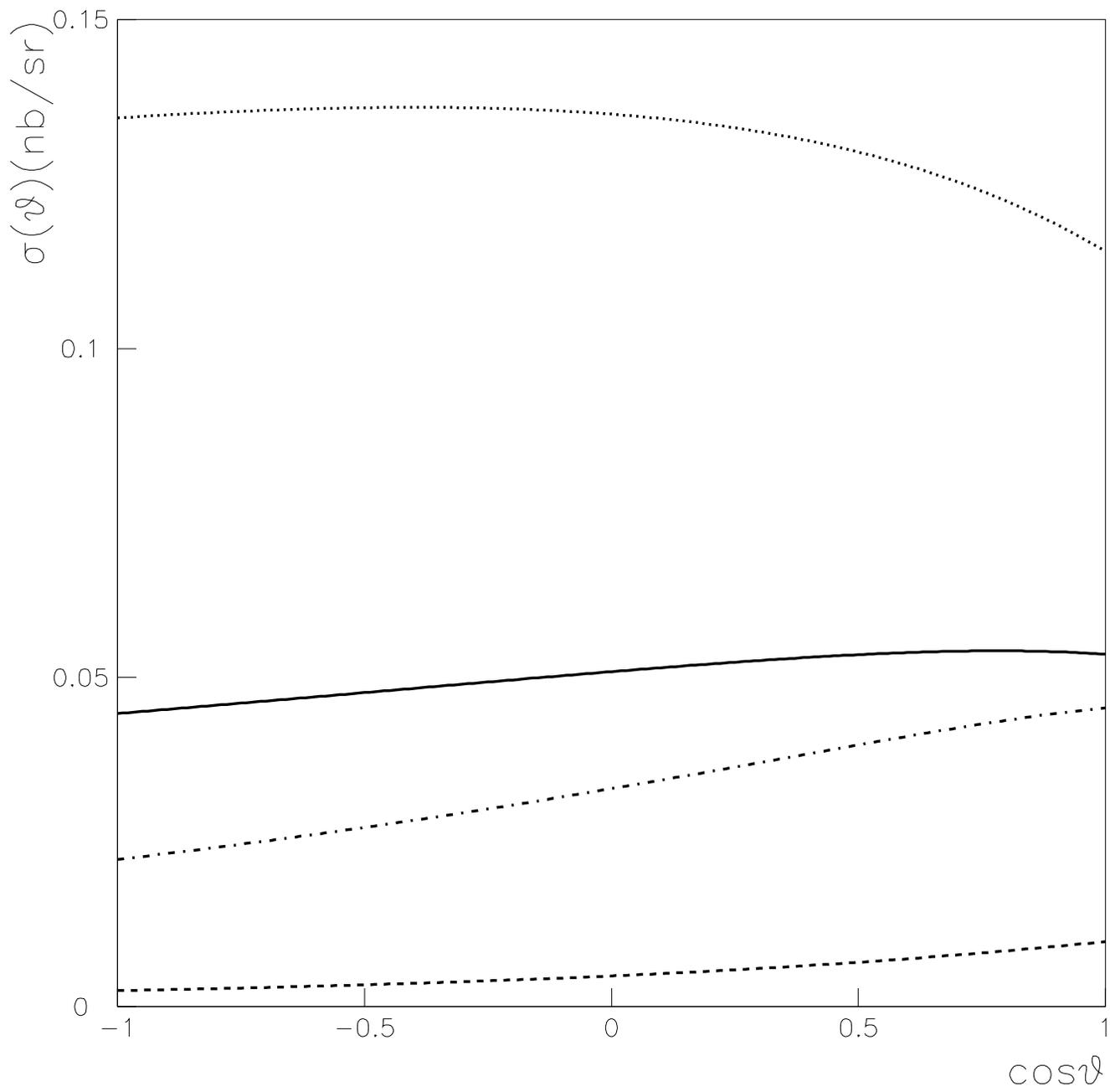
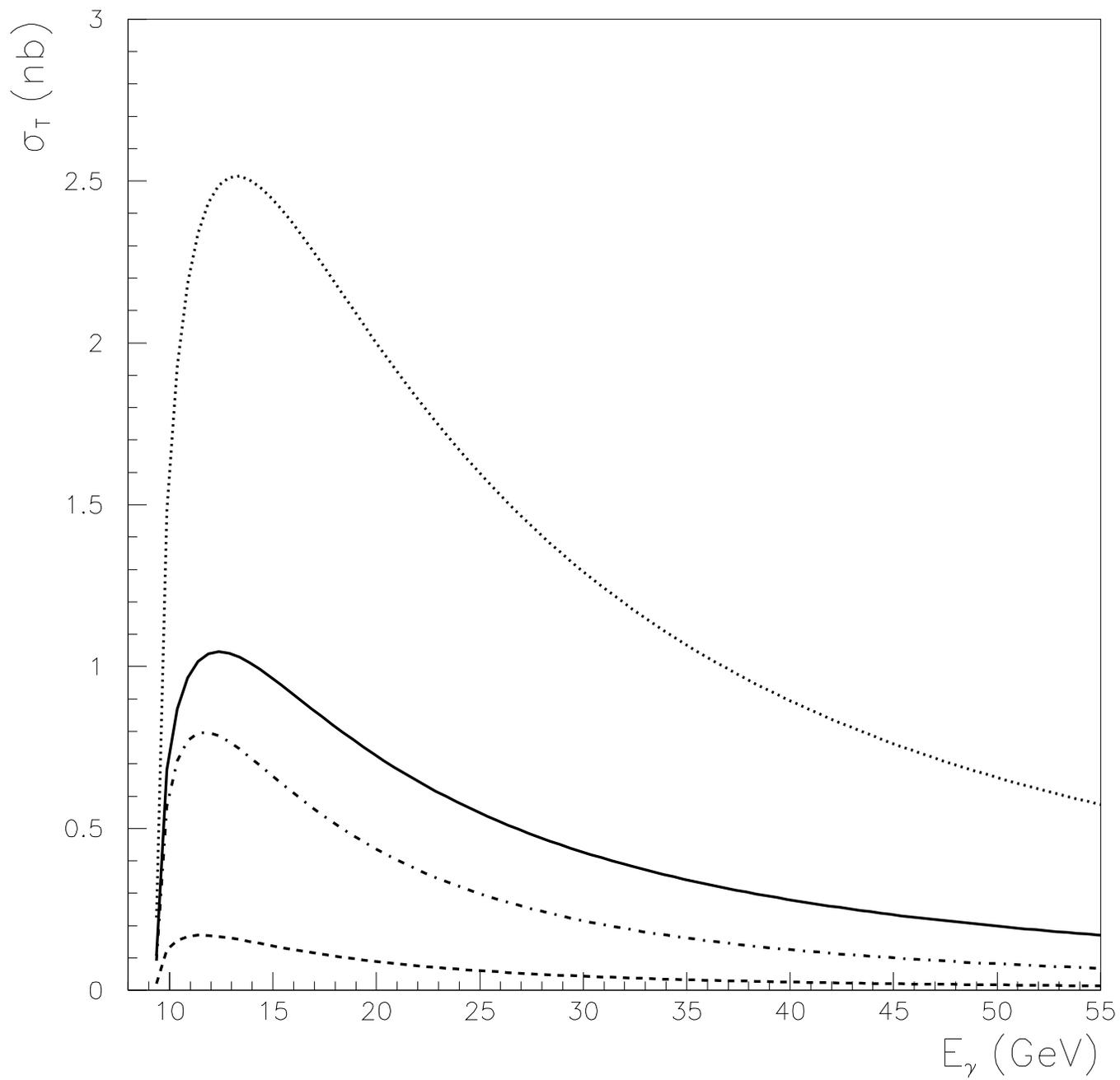
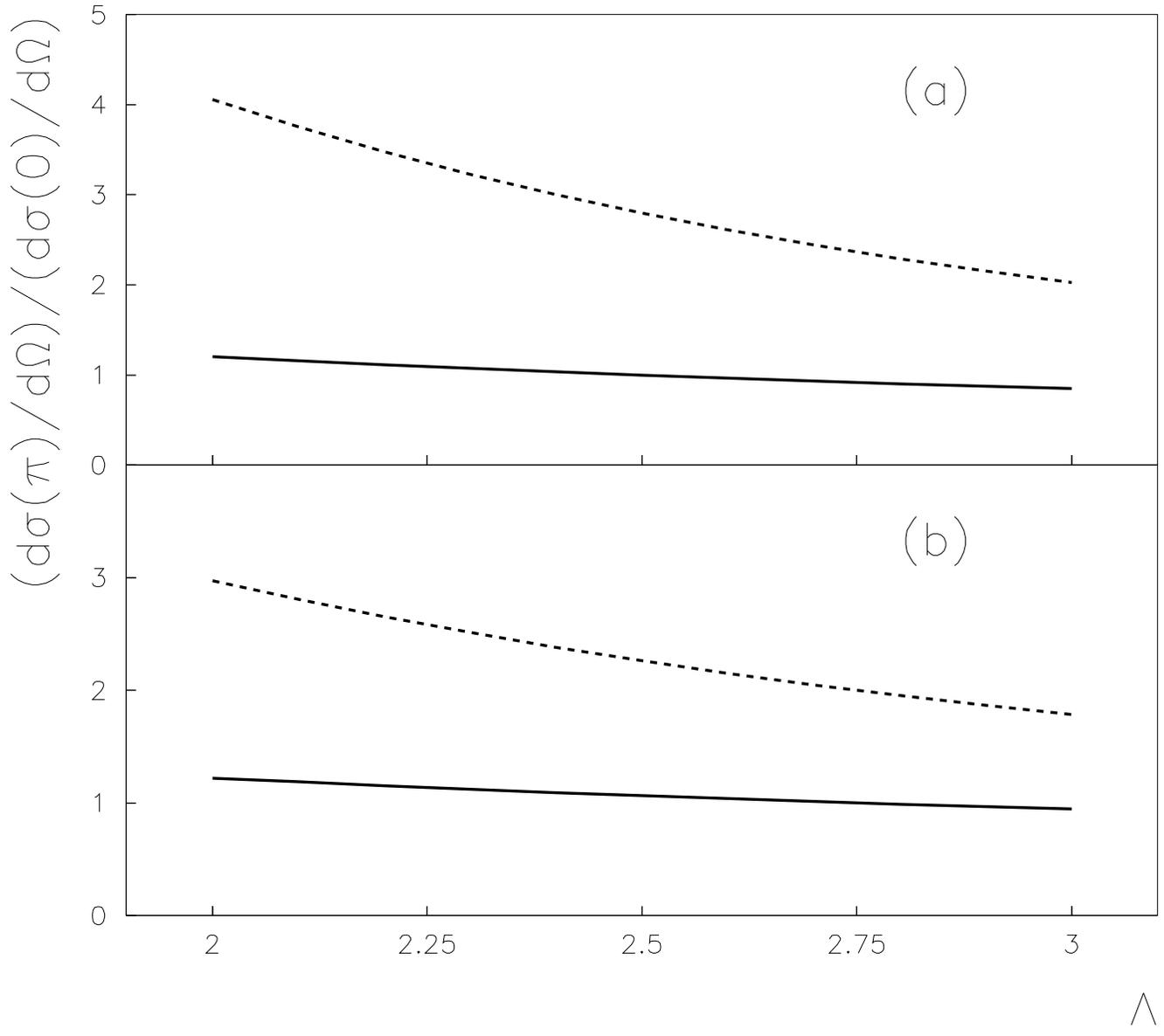

The energy dependence of the total cross section is reported in Fig. \ref{fig:exc} for $\gamma+N\to  \Lambda_c^+ +\overline{D^*}$.
We see that this cross section sharply increases at threshold  and has a maximum around 10 GeV, which position slightly depends on the choice of the form factor.
The sensitivity of the backward-forward asymmetry to $\Lambda$ is shown in Fig. \ref{fig:ratio}, for $n=1$ and 2.

Note that in the framework of the considered model, for $\gamma+N\to {\cal B}_c+ \overline{D^*}$, the absolute value of the cross section  is determined by the product $g_{D^*D\gamma}g_{N{\cal B}_cD}$ of the electromagnetic and strong coupling constants,
whereas the shape of the $t-$dependence of the differential cross section,  
and the energy behavior of the total cros section are driven only by the parameters of the phenomenological form factors, $n$ and $\Lambda$, which can therefore be determined from the experimental data, when available. Having determined the correct form factor, the product of the coupling constants  $g_{D^*D\gamma}g_{N{\cal B}_cD}$ can be derived from the absolute value of the differential cross section (for a fixed value of $\cos\theta$ and 
$E_{\gamma}$). Finally the ratio of cross sections on proton and neutron targets will help to determine the electromagnetic coupling constants:
\begin{equation}
R=d\sigma(\gamma n\to \overline{D^{*-}}\Lambda_c^+)/d\sigma(\gamma p\to \overline{D^{*0}}\Lambda_c^+)=g^2_{D^{*-}D^-\gamma}/g^2_{D^{*0}D^0\gamma}.
\label{eq:ratio}
\end{equation}

The results of the SLAC experiment \cite{Abe}, concerning open charm photoproduction at $E_\gamma$=20 GeV, constrain the values of the coupling constants and the choice of the form factor $F(t)$. At this energy, it is found that the probability of $D^*$-production per charm event is $0.17\pm0.11$. The total cross section for charm photoproduction being $56^{+24}_{-23}$ nb, the  upper limit for the cross section of $\gamma+p\to \Lambda_c^+ +\overline{D^{*0}}$ can be 
estimated to be of the order of 10 nb.

So, from our calculation (see Fig. \ref{fig:exc}) we can estimate the product of the strong and electromagnetic coupling constants:
$R=g^2_{D^{*0}D^0\gamma} g^2_{N\Lambda_cD}$, for the different choices of the phenomenological form factor $F(t)$:
$R_{exp}\le$ 5 (for n=1 and $\Lambda =3$ GeV), 15 
(for n=1 and $\Lambda =2$ GeV), 20 
(for n=2 and $\Lambda =3$ GeV), and 100 
(for n=2 and $\Lambda =2$ GeV). These numbers may be compared with some conservative theoretical estimates for the value $R$. As a starting point let us consider $SU(4)$-symmetry, which predicts the same value for the coupling constants $g_{N\Lambda_cD}$ and $g_{N\Lambda K}$ \cite{Li00}. This last coupling constant has been estimated, from the analysis of processes of photoproduction of strange particles on nucleon \cite{Ad90}, to be in the interval $13.2\le |g_{N\Lambda K}|\le 15.7$. QCD sum rules \cite{Ch96} predict a smaller value for this constant, $g_{N\Lambda K}=6.7\pm 2.1$.

Taking into account that theoretical predictions [2-12] are consistent with the value  of the radiative width, $\Gamma (D^*\to D\gamma )\ge $10 keV, one can deduced a 'theoretical' value for the ratio $R$, $R_{th}\ge$ 1300, i.e.
$R_{th}\gg R_{exp}$, for any considered parametrization of the form factor $F(t)$. In particular the value $n=1$ should be excluded; for $n=2$ the smallest value of the cut-off parameter $\Lambda$ seems preferable. But even for $n=2$ and $\Lambda$ = 2 GeV, we find $g_{N\Lambda_cD}<1$, in large disagreement with $SU(4)$ expectations. Note that most calculations for processes like $J/\psi +N\to \Lambda_c+D$ (in connection with the interpretation of quark gluon plasma), rely on $SU(4)$-symmetry \cite{Si01}. Taking the form factor with $n=2$ and $\Lambda$ = 2 GeV, we predict a maximal value of the cross section for 
$\gamma+p\to  \Lambda_c +\overline{D^{*0}}$
of 20 nb, in the near threshold region. The ugraded JLab, with a tagged photon beam of luminosity ${\cal L} =10^{-29}\div 10^{-30}$ cm$^{-2}$s$^{-1}$, can, in principle, investigate such processes.

In order to test the proposed model, polarization phenomena in 
$\gamma+p\to  \Lambda_c +\overline{D^*}$ will be very useful. Polarization effects can be easily predicted. For example, the beam asymmetry, $\Sigma$, vanishes, for any choice of the phenomenological form factors and for any values of the coupling constants, independently on kinematical conditions. The polarization of the produced $D^*$ is characterized by a single non-zero element of the density matrix, $\rho_{11}=1/2$, with a $\sin^2\theta_D$-distribution for  $D^* \to D +\pi$,  where $\theta_D$ is the angle between the three-momenta of $D$ and the initial $\gamma$, in the $\overline{D^*}$ rest frame system. Due to the fact that the $D-$exchange amplitudes are real, all T-odd polarization effects are identically zero. However these observables do not vanish, if one takes into account the strong final 
$\overline{D^*}\Lambda_c$-interaction, which might play a role near the reaction threshold.
Another source of T-odd effects is the unitarity condition in the $s$-channel, due to the large number of possible intermediate states in $\gamma+p\to  \Lambda_c +\overline{D^*}$, even in the threshold region.
%\section{Conclusions}

In conclusion we have calculated the differential and total cross sections for the associative charm photoproduction of vector $\overline{D^*}$, through a pseudoscalar $D-$meson exchange model, in analogy with meson and strange particle photoproduction. We introduced phenomenological form factors, with a larger cut-off parameter in comparison with the usual value for $\rho,\omega$ vector mesons. We found large sensitivity of the differential and total cross section for $\gamma+p\to  \Lambda_c +\overline{D^*}$ to the value of the cut-off parameter. Finally we found sizeable values for the cross section in the threshold region and predicted the energy dependence for these reactions, which  will be  experimentally accessible in next future. The existing experimental data on charm photoproduction at $E_{\gamma}$=20 GeV allow to constrain the value of the $g_{N{\cal B}_cD}$ coupling constant.

\begin{figure}
%\mbox{\epsfxsize=15.cm\leavevmode\epsffile{fig1.ps}}
\caption{Diagram for $D^*_c$-photoproduction, through $D_c$-exchange}
\label{fig:dst}
\end{figure}

\begin{figure}
%\mbox{\epsfxsize=15.cm\leavevmode \epsffile{fig2.ps}}
%\vspace*{-2.5 true cm}
\caption{$\cos \vartheta$-dependence of the differential cross 
section  $\sigma(\vartheta)\equiv 
(d\sigma/d\Omega)/(g^2_{N{\cal B}_cD}g^2_{D^* D \gamma})$ for the reaction $\gamma+p\to  {\Lambda}_c^+ +\overline{D^{*0}}$, at $E_{\gamma}=11$ GeV. Different curves correspond to different values of $n$ and $\Lambda$:  $n=1$ and $\Lambda$=2 (solid line), $n=2$ and $\Lambda$=2 (dashed line), $n=1$ and $\Lambda$=3 (dotted line), $n=2$ and $\Lambda$=3 (dash-dotted line).
}  
\label{fig:difcs}
\end{figure}

\begin{figure}
%\mbox{\epsfxsize=15.cm\leavevmode \epsffile{fig3.ps}}
\caption{Energy dependence of the total cross section $\sigma_T=\int \sigma(\vartheta)d\Omega$, for the reaction 
$\gamma+p\to  {\Lambda}_c^+ +\overline{D^{*0}}$. Notations as in Fig. \protect\ref{fig:difcs}.}
\label{fig:exc}
\end{figure}
\end{document}